\documentclass[printer, bibyear]{aa}

\usepackage[varg]{txfonts}
\usepackage{graphicx}
\usepackage{setspace}
\usepackage{tocloft}
\usepackage[normalem]{ulem}

\begin{document}

\title{A Combined Compton and Coded-aperture Telescope for Medium-energy Gamma-ray Astrophysics}

\author{Michelle Galloway,\inst{1} 
\and Andreas Zoglauer,\inst{2} 
\and Steven E. Boggs,\inst{2} 
\and Mark Amman\inst{3}}

\institute{Universit\"{a}t Z\"{u}rich, Physik-Institut, Winterthurerstrasse 190, Z\"{u}rich, Switzerland, CH-8057
\and University of California at Berkeley, Space Sciences Laboratory, 7 Gauss Way, Berkeley, California, USA, 94720
\and Lawrence Berkeley National Laboratory, 1 Cyclotron Road, Berkeley, California, USA, 94720}

\date{Received date / Accepted date }


\abstract{A future mission in medium-energy gamma-ray astrophysics would allow for many scientific advancements, e.g. a possible explanation for the excess positron emission from the Galactic Center, a better understanding of nucleosynthesis and explosion mechanisms in Type Ia supernovae, and a look at the physical forces at play in compact objects such as black holes and neutron stars. Additionally, further observation in this energy regime would significantly extend the search parameter space for low-mass dark matter. In order to achieve these objectives, an instrument with good energy resolution, good angular resolution, and high sensitivity is required. In this paper we present the design and simulation of a Compton telescope consisting of cubic-centimeter Cadmium Zinc Telluride (CdZnTe) detectors as absorbers behind a silicon tracker with the addition of a passive coded mask. The goal of the design was to create a very sensitive instrument that is capable of high angular resolution. The simulated telescope showed achievable energy resolutions of 1.68$\%$ FWHM at 511 keV and 1.11$\%$ at 1809 keV, on-axis angular resolutions in Compton mode of 2.63$^{\circ}$ FWHM at 511 keV and 1.30$^{\circ}$ FWHM at 1809 keV, and is capable of resolving sources to at least 0.2$^{\circ}$ at lower energies with the use of the coded mask. An initial assessment of the instrument in Compton imaging mode yields an effective area of 183 cm$^{2}$ at 511 keV and an anticipated all-sky sensitivity of \mbox{3.6 x 10$^{-6}$ photons cm$^{-2}$ s$^{-1}$} for a broadened 511 keV source over a 2-year observation time. Additionally, combining a coded mask with a Compton imager to improve point source localization for positron detection has been demonstrated.}


\maketitle 


\section{Introduction}
\label{sect:intro} 
The highly penetrative nature of gamma rays allows for a unique probe into the most violent explosions and dynamic sources in the Universe. As they travel through the Galaxy relatively undeterred, detected photons in the MeV to GeV regime point directly back to their cosmic source. Obtaining spatial, spectral, and temporal information at these wavelengths will help us to decipher the physical mechanisms behind astrophysical sources and events, to discriminate between different theoretical models, and may reveal new sources and phenomena. Particularly, imaging in this regime provides valuable information regarding the type, morphology, and spatial distribution of gamma-ray sources.

Previous missions using imaging telescopes such as COMPTEL (Schoenfelder et al. 1993) and the Compton Spectrometer and Imager (COSI) instrument (Kierans 2017) have demonstrated good Compton imaging capabilities in the medium-energy gamma-ray regime, for example, the all-sky map of $^{26}$Al by COMPTEL (Oberlack 1996) and the image of the Crab Nebula by NCT (Bandstra 2011). Additionally, coded-aperture imagers such as SPI (SPectrometer on INTEGRAL) onboard the INTEGRAL (INTErnational Gamma-Ray Astrophysics Laboratory) spacecraft (Vedrenne 2003) have provided images of the Galactic Plane and a Galactic all-sky map of the 511 keV positron annihilation line (Bouchet 2010).

A next generation telescope in the medium-energy gamma-ray regime requires good energy resolution for the study of line emissions resulting from Galactic nucleosynthesis. Good angular resolution is needed to resolve point sources, thus providing a means to correlate gamma-ray emission with sources that are seen in other wavebands. Additionally, a telescope must have good efficiency and a high sensitivity to detect and image sources above a complex background. The high sensitivity needed for an astrophysical instrument can be achieved through the capabilities of a Compton telescope with regards to its background rejection capabilities and as a result of excellent energy, position, and timing resolution. Furthermore, adding a coded aperture to a Compton telescope can significantly improve its angular resolution to well below the Doppler limit (Zoglauer 2003) within its masked field-of-view. Coded-aperture Compton telescopes (CACTs) have previously been investigated with the IBIS instrument onboard the INTEGRAL satellite (Forot 2007)

This paper describes a preliminary study of a Compton imager combined with a passive coded mask with the goal of achieving high sensitivity as well as the ability to resolve point sources in crowded fields like the Galactic Center Region. The Compton imager utilizes cubic-centimeter cadmium zinc telluride (CdZnTe) detectors as absorber planes behind an array of double-sided silicon strip detectors (DSSD Si) as particle tracker. Large-volume CdZnTe detectors are an attractive candidate for a future instrument because of their good absorption, simple design, and minimal or no cooling requirements. A passive mask covering a narrow (10$^{\circ}$) field-of-view (FoV) is added to the Compton telescope in order to improve the on-axis angular resolution for dedicated observations. 

Through simulations of sources in the energy range between 200 keV and 6 MeV with particular emphasis on characterizing the instrument at 511 keV (positron annihilation line) as well as at 1809 keV ($^{26}$Al line), an initial assessment of the achievable energy resolution, angular resolution, and effective area is given. The sensitivity for line sources and a continuum source is estimated based upon the expected background in a low-Earth orbit. Additionally, localization of two 511 keV point sources separated by 0.2$^{\circ}$ is demonstrated using combined Compton and coded-aperture imaging modes.


\section {Scientific Motivation}
The MeV to GeV energy range remains largely unexplored since the success of COMPTEL more than two decades ago, yet many open questions remain that can only be fully answered through observations in this regime. For example, excesses of both extragalactic, in the $\sim$0.2-100 MeV regime (Ajello 2009), and Galactic Center (GC) gamma rays, the latter particularly from positrons (Prantzos 2011), were consistently observed over many decades without clear indications of their origins. The GC excess has no distinguishable counterparts at other wavelengths, and it is not yet determined if the emission is diffuse or from point sources. Several candidates have been proposed as explanation, such as cosmic rays, low-mass dark matter, or millisecond pulsars. In the case of the extragalactic MeV background, its observed intensity cannot be accounted for by any known source population or emission mechanism. Future observations in the MeV domain, particularly with high angular resolution, would help to distinguish between diffuse emission candidates and point sources of these excesses. Spectral information and determination of the redshift and luminosity functions would provide additional clarification on the nature of the MeV background. Further, pointed observations of e.g. the Galactic Center region or Dwarf spheroidal galaxies in this energy regime will significantly widen the search for annihilation products from MeV to \mbox{sub-GeV} dark matter, thus providing the capability to either detect a signature or set more stringent constraints on existing models (Boddy 2015).

Another area where exploring the MeV domain can yield significant advances is in the understanding of compact objects, e.g. Active Galactic Nuclei (AGN), core-collapse and thermonuclear supernovae, and neutron stars. For relativistic jets and outflows from AGNs, the transition between the low-energy (X-ray) continuum to the spectral shapes observed in the GeV to TeV range occurs in the MeV regime. Therefore this gap holds a wealth of information regarding the acceleration processes that occur within and around AGNs (Padovani 2017). For Type 1a supernovae, spectral and luminosity measurements in the MeV range can provide important information about their evolution with respect to look-back time and metallicity; such an understanding is a prerequisite for their use as standard candles in precision cosmology (Phillips 1993). With respect to neutron stars, a recent example of the relevance of gamma-ray astronomy at the dawn of a multi-messenger era is the detection of the gravitational wave GW170817 in temporal coincidence with the GRB 170817 (Goldstein 2017). Although it was long suspected that a neutron star merger could be a progenitor of a short-duration gamma-ray burst, confirmation was only possible through such a coincident measurement. As additional information in both the prompt and afterglow emissions from GRBs is encoded in the MeV domain, future detection and imaging in this regime will contribute towards our understanding of the properties, environment, and evolution of GRB sources as well as provide the multi-messenger community with relatively instantaneous localization of progenitor systems.


\section{Telescope Design}
\begin{figure*}[ht]
\centering
\includegraphics[width=0.55\linewidth]{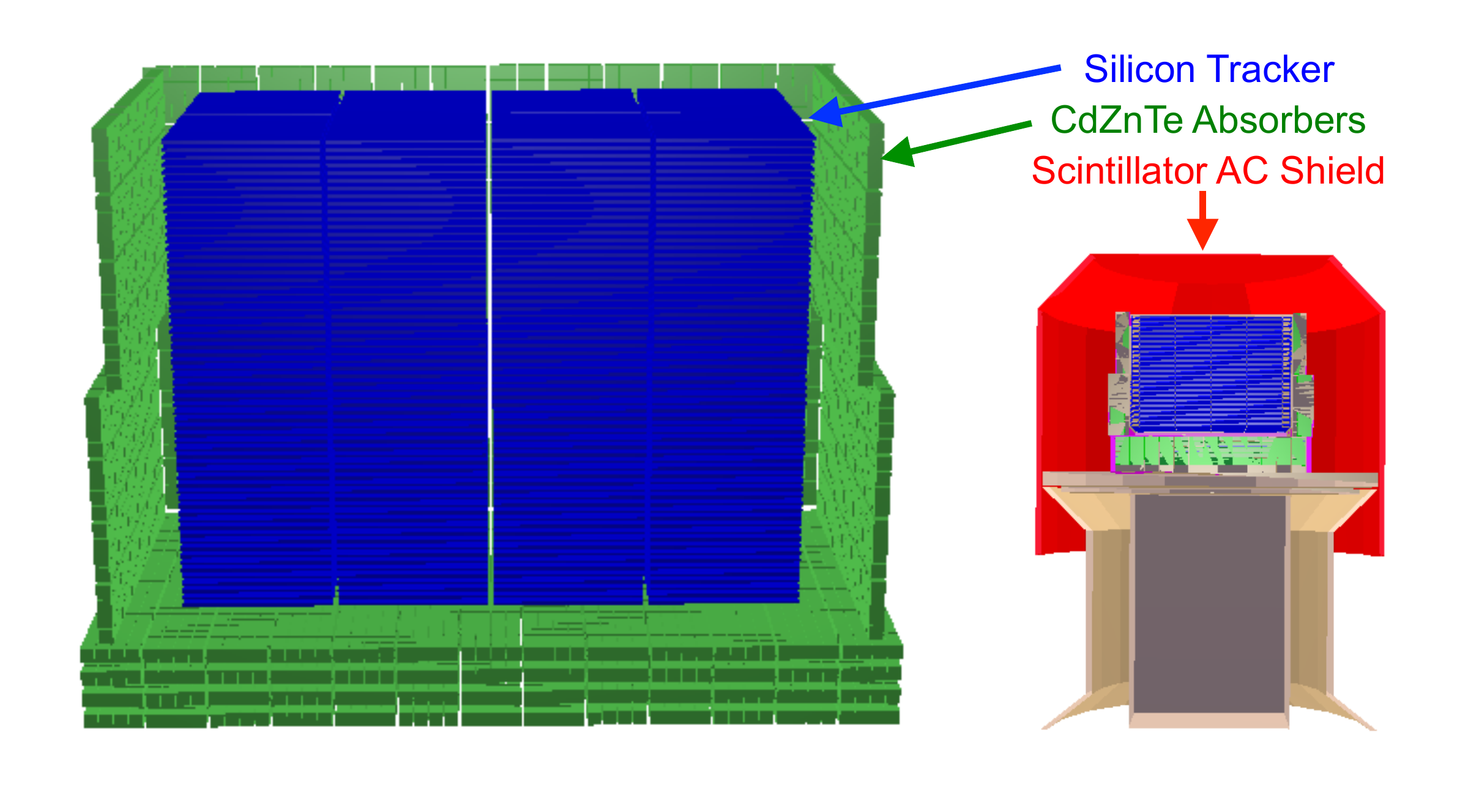}
\caption
{\label{fig:CoCo_wholemodel} Left: Mass model of Compton telescope consisting of a silicon tracker array (blue) and CdZnTe absorber planes (green).  Right: Mass model of organic scintillator \mbox{anti-coincidence} (AC) shield surrounding tracker and absorber detectors.}
\end{figure*}

\begin{figure*}[ht]
\centering
\includegraphics[width=0.70\linewidth]{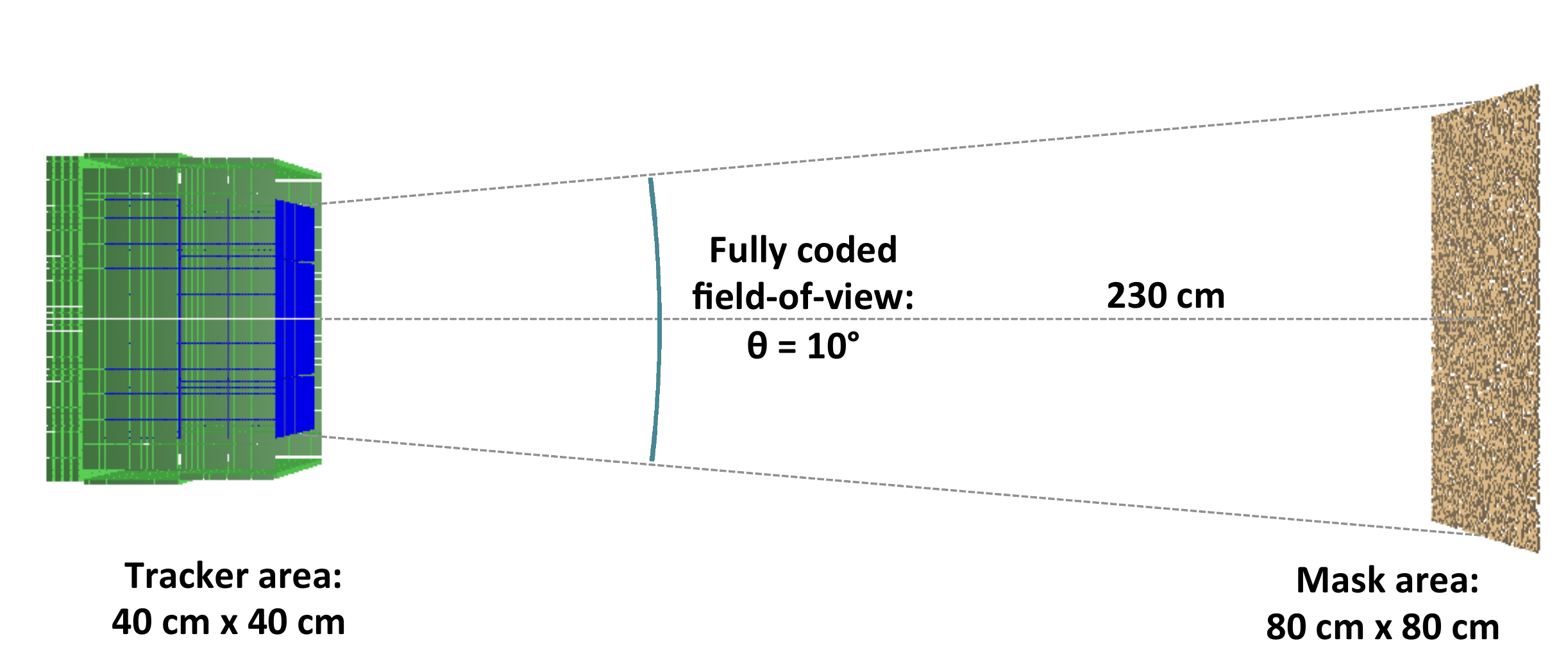}
\caption
{\label{fig:CoCo_modelwithmask}The Compton telescope combined with the tungsten coded mask with a separation distance of 2.3 meters, allowing for a fully-coded field-of-view of 10$^{\circ}$. }
\end{figure*}

The proposed design of the telescope is shown in Fig.~\ref{fig:CoCo_wholemodel}. The Compton telescope, left, consists of a silicon tracker (blue) in which the first Compton interaction takes place and a CdZnTe absorber (green) surrounding the tracker on 5 sides. Figure~\ref{fig:CoCo_wholemodel}, right, shows the Compton telescope mounted on a support structure and surrounded by an organic plastic scintillator dome (red). The dome serves as an \mbox{anti-coincidence} (AC) shield to veto charged particle interactions. An electronics box is contained within the pedestal of the support structure. The passive coded mask comprised of tungsten pixels above the tracker is shown in Fig.~\ref{fig:CoCo_modelwithmask}. Additional passive materials such as front-end electronics for all detectors, housing for the CdZnTe detectors, and a support frame for the tracker are included in the model. The total mass of the proposed instrument with support structure and shielding is $\sim$530 kg. This mass is feasible for a Mid-sized Explorer (MIDEX) NASA mission.

\subsection{Silicon Tracker}
The Compton tracker consists of multiple layers of double-sided silicon strip detectors and is based upon the Gamma-Ray burst Investigation via Polarimetry and Spectroscopy (GRIPS) telescope (Greiner 2008; Zoglauer 2008). A low-Z material such as silicon (Z = 14) is a good choice for a Compton scatter plane, as an incident photon can scatter many times in the tracker before becoming fully absorbed. These multiple interactions allow for a higher probability of correct Compton reconstruction, thus improved background rejection. Additionally, low-Z materials such as silicon have a smaller contribution from Doppler broadening to the angular resolution as well as a higher Compton cross-section (relative to its photoabsorption cross-section) as compared to medium-Z to high-Z materials such as Ge and CdZnTe. Simulated detector parameters include low-energy and trigger thresholds of 5 keV and 10 keV, respectively, and a one sigma FWHM energy resolution over a 0.2 to 10 MeV energy range. This allows for a telescope with good overall energy resolution.

Each silicon wafer has a 10 by 10 cm wide active area surrounded by a 1.5 mm guard ring. On the top and bottom of the silicon material are 0.5 mm wide strip electrodes. The top strips are orthogonal to the bottom strips in order to obtain x-y position information for each interaction. The spatial resolution of the silicon detectors is given by the 0.5 mm strip pitch, thus is a contributing factor that allows for good angular resolution of the telescope. Several wafer thicknesses were simulated to optimize the angular resolution and effective area at 511 keV before arriving at a thickness of 2 mm.

The tracker consists of 4 by 4 wafers, yielding a total geometric surface area of 1600 cm$^{2}$. As with the GRIPS telescope, 64 layers of wafers were used (1,024 total Si detectors) with the vertical spacing between layers set to 0.5 cm in order to provide a high Compton efficiency over a wide FoV. This yields a total tracker depth of $\sim$33 cm. The total mass of silicon in the tracker is $\sim$50 kg.

\subsection{CdZnTe Absorbers}
For this work, the lanthanum bromide calorimeters that were used in the GRIPS design were replaced with 1-cm$^{3}$ coplanar-grid CdZnTe detectors (Luke 1995, 1996; Amman $\&$ Luke 1999). The large volume, ambient-temperature CdZnTe detectors have the advantages of simplified electronics, low power consumption, a high active to passive material ratio, and a high average Z (Z$_{ave}$ = 49), making them a good candidate for a Compton telescope absorption plane. Particularly, having little or no cooling requirement for the detectors allows for reduced background due to minimization of passive material surrounding the instrument. The CdZnTe detector response is well understood after successful benchmarking of laboratory measurements with the High Efficiency Multimode Imager (HEMI) (Galloway 2011), a combined Compton and coded-aperture instrument developed for nuclear security applications. As a result of detector development during the HEMI campaign, the cubic-centimeter CdZnTe detectors were fabricated with good spectral performance (Amman 2009). The energy resolution of the highest performing CdZnTe detector was simulated in this case, i.e.\ 1.5$\%$ FWHM at 662 keV, as a reasonable approximation of the obtainable energy resolution over a long-term production phase. Although this assumes room temperature operation, it is noted that a $\sim$30$\%$ improvement in energy resolution can be achieved with moderate cooling to -20$^{\circ}$C (Amman 2006). A benchmarked time resolution of 0.5 $\mu$s was used in the simulation.

The absorber planes consist of a single layer of 7200 elements surrounding the tracker on 4 sides and a 52 by 52 array of elements 4 layers deep underneath the tracker (10,816 elements). Each individual element includes front-end electronics and a Lexan housing, as in the HEMI design. The number of layers surrounding the tracker on all 5 sides was optimized to achieve full absorption of high-energy photons within the sensitive volume of the telescope without having excess detector material. The final model contains 18,016 CdZnTe detector elements with a total mass of $\sim$103 kg.

\subsection{Coded Mask}
The coded aperture design began with the choice of a mask pattern with a specified ratio of transparent elements to total mask area, i.e. the open fraction denoted as \textit{f$_{e}$}. To obtain the best signal-to-noise ratio, the open fraction of the mask was optimized according to the expected point source intensity relative to the background intensity (Gunson $\&$ Polychronopulos 1976). It was found that the optimum fraction is \mbox{\textit{f$_{e}$} = 1/2} if the background level dominates over the source intensity. This result was reviewed under various detection scenarios (in't Zand 1992; Skinner 2008). Considering the real-world application in medium-energy gamma-ray astrophysics, an open fraction of \mbox{\textit{f$_{e}$} = 1/2} is an appropriate choice, as used by the Burst Area Telescope (BAT) on Swift (Barthelmy 2005).

The mask with telescope is shown in Fig.~\ref{fig:CoCo_modelwithmask}. The mask element spacing, the distance from the mask to the Compton telescope, and the geometric area of the mask were designed to provide an angular resolution, \textit{$\Theta$}, in pointing mode of  0.125$^{\circ}$ \mbox{($\sim$7.5 arcminutes)} and a 10$^{\circ}$ FoV. The mask consists of cubic tungsten pixels (Z = 84) with a size of 80 x 80 x 0.5 cm arranged in a 160 x 160 random mask pattern (not optimized). A support structure for the mask is not included in the model. The mask-detector distance is 230 cm. The geometric mask area is \mbox{6400 cm$^{2}$} with a total tungsten mass of $\sim$30 kg.

\subsection{Anti-coincidence Shield}
An organic plastic scintillator anti-coincidence (AC) shield surrounds the Compton telescope on all sides, (Fig.~\ref{fig:CoCo_wholemodel}, red). The AC shield allows for the veto of external background events induced primarily by Albedo and semi-trapped charged particles, as well as from cosmic rays that are not already reduced by a low-Earth orbit (described in Section 6.1). An energy resolution of 10 keV (1$\sigma$ Gaussian) and a trigger threshold of 50 keV were used for the shield. The total mass of the plastic scintillator and related components is $\sim$70 kg.

%
\section{Analysis Tools and Methods}
The Medium Energy Gamma-ray Astronomy library software package, MEGAlib, (Zoglauer 2006) was used for all aspects of the data analysis including simulations, Compton event reconstruction, and both Compton and coded-aperture imaging. MEGAlib is written in C++ and based on ROOT (Brun et al.1997). It is open source software that was designed for use with gamma-ray detectors, particularly for imaging with Compton telescopes. 

Monte Carlo simulations were performed using Cosima (Zoglauer 2009), the simulation toolkit contained in MEGAlib. In order to simulate the performance of the telescope, first a realistic geometry was created using the Geomega package in MEGAlib (Fig.~\ref{fig:CoCo_wholemodel} and~\ref{fig:CoCo_modelwithmask}). The geometry specifies the dimensions of each volume, the elemental composition, and density of all passive and active volumes within the detector and surrounding environment. The simulator imports the geometry and converts it into a Geant4 (Agostinelli 2003) format. Cosima then uses the specified source information to simulate particle propagation and interactions with the modeled materials, e.g., photoabsorption, Compton and Rayleigh scattering. The resulting Cosima output file contains the simulated interaction information including the time, type, position, and energy deposition of each interaction. The simulated hits are convolved with the detector response using the detector effects engine contained in MEGAlib.

\subsection{Compton Reconstruction}
For Compton reconstruction as applied to this work, coincident hits are merged together in the simulation, therefore adjacent hits were clustered together to form Compton events. Compton Sequence Reconstruction, CSR, (Zoglauer 2005) is then applied to the convolved data to determine the sequence of each event, thus finding the initial photon energy and the most likely scattering angle of the incoming photon. It was required in CSR that the first hit occur in the silicon tracker. Nominally this implies a FoV of $<$90$^{\circ}$ due to the positioning of the Si tracker inside the CdZnTe well, however due to passive material surrounding each CdZnTe detector, transmission still occurs at higher incidence angles to allow for the first interaction to take place in the tracker. Therefore the performance was evaluated over a 120$^{\circ}$ FoV. Additionally, an Earth horizon cut of 90$^{\circ}$ was used to eliminate events which originate at or below the horizon to reduce background events generated from albedo photons.

This list of reconstructed events then undergo imaging and other high-level analysis such as application of event cuts, estimation of angular resolution measurement (ARM), and an assessment of photon interaction locations and energies, using the Mimrec tool in MEGAlib (Zoglauer 2011). Imaging algorithms are used to generate a Compton backprojected image within a user-specified coordinate system. Additionally, a List-Mode Maximum Likelihood Expectation Maximization (ML-EM) algorithm can be applied to further refine the image (Wilderman 1998).

\subsection{Combined Compton and Coded-aperture Imaging}

\begin{figure}[ht]
\begin{center}
\begin{tabular}{c}
\includegraphics*[width=0.95\linewidth]{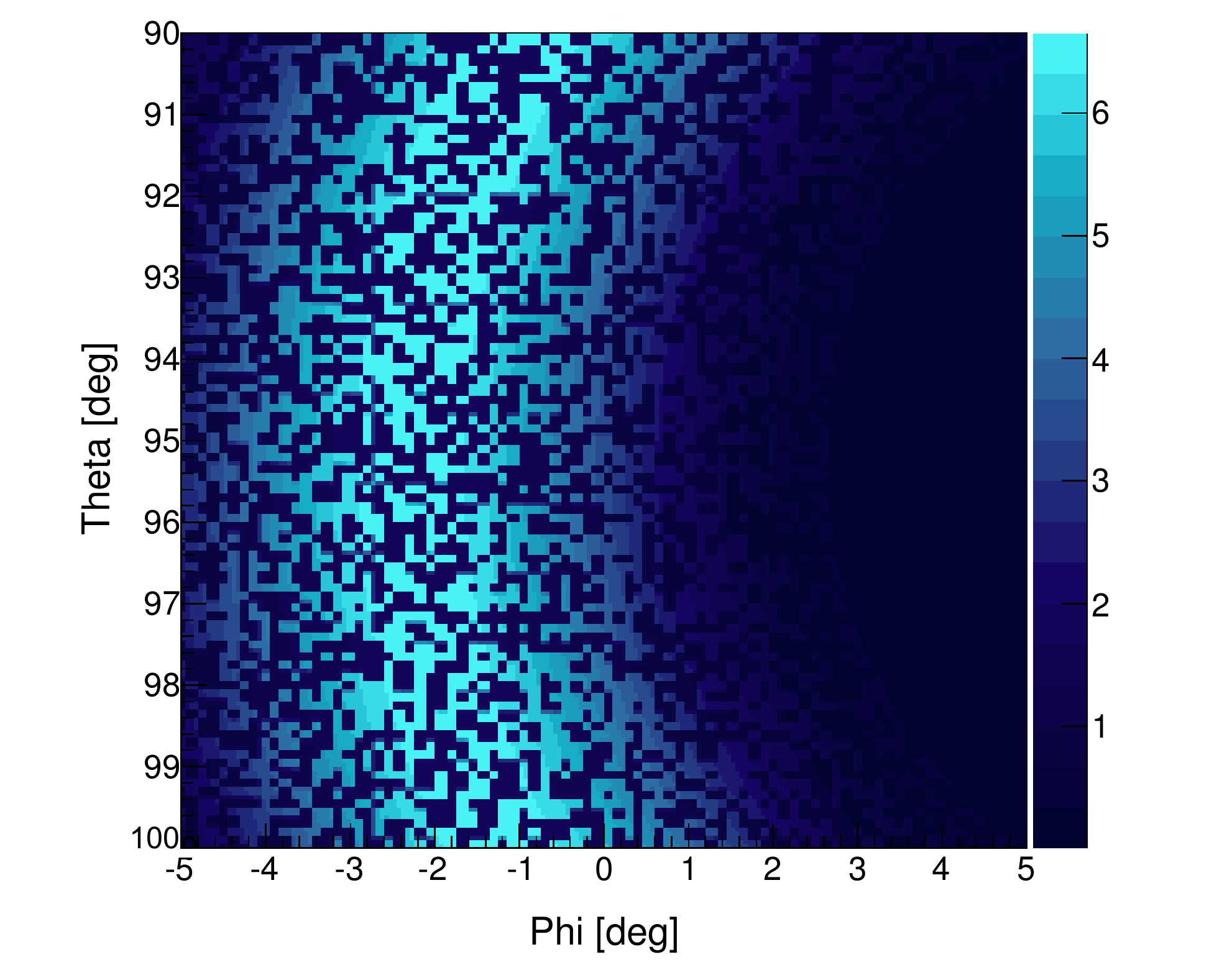}
\end{tabular}
\end{center}
\caption
{\label{fig:oneevent}Reconstructed Compton circle (backprojection) overlaid with the projected mask pattern. The pixellation of the mask within the 10$^{\circ}$ FoV significantly improves the angular resolution.}
\end{figure}

In combined-imaging mode, first Compton event reconstruction and imaging is performed. The region of overlap of the Compton circles identifies the source location with an angular resolution as given by the ARM distribution. The coded-aperture imaging technique is then applied to these selected Compton events, where the first hit can occur in any layer of the silicon tracker. For each reconstructed point of origin, the probability that the photon was photoabsorbed as it passed through the mask is calculated. The result is a shadow pattern projected onto the Compton reconstructed point of origin. 

The strength of adding a coded mask to a Compton telescope is exemplified in Fig.~\ref{fig:oneevent} which shows a coded-mask pattern backprojected onto a reconstructed Compton cone from a single event. The pixellation of the pattern limits the point of origin from the width of the ARM distribution to that given by the mask geometry, thus greatly improving the angular resolution. Likewise, the background rejection from the Compton reconstruction eliminates a significant area of the backprojected mask, thus coding noise is not an issue and greatly improved source localization is obtained as compared to coded-aperture mode alone.

\section{Simulated Performance}
The instrument was placed in a low-Earth, near-equatorial orbit. A series of monoenergetic far-field line sources from 200 keV to 10 MeV were simulated on-axis, as described in Table~\ref{tab:spaceHEMI_lines}. A 511 keV source broadened by $\sigma$ = 1.25 keV was also simulated at various incident angles. The broadening is based upon observations of the line from INTEGRAL/SPI (Jean 2006). Additionally, a Crab-like source with a power law index of 2.17, also observed by SPI (Sizun 2004, Jourdain and Roques 2009), was simulated to predict the sensitivity of the telescope to continuum sources.

\begin{table}[ht]
\label{tab:spaceHEMI_lines}
\begin{center}       
\begin{tabular}{lllll}
\hline\hline
\rule[-1ex]{0pt}{3.5ex}    Energy  (keV) &      Source    \\
\hline
\rule[-1ex]{0pt}{3.5ex} 200 & nonspecific	\\
\hline
\rule[-1ex]{0pt}{3.5ex} 511 & e$^{+}$e$^{-}$, $\beta$$^{-}$ decays, etc. \\
\hline
\rule[-1ex]{0pt}{3.5ex} 847 & $^{56}$Ni $\rightarrow$ $^{56}$Co  \\
\hline
\rule[-1ex]{0pt}{3.5ex} 1157 & $^{44}$Ti  \\
\hline
\rule[-1ex]{0pt}{3.5ex} 1332 & $^{60}$Fe $\rightarrow$ $^{60}$Co  \\
\hline
\rule[-1ex]{0pt}{3.5ex} 1809 & $^{26}$Al  \\
\hline
\rule[-1ex]{0pt}{3.5ex} 4440 & $^{12}$C$^{*}$ molecular deexcitation  \\
\hline
\rule[-1ex]{0pt}{3.5ex} 6130 & $^{16}$O$^{*}$ molecular deexcitation  \\
\hline
\rule[-1ex]{0pt}{3.5ex} 511 $\pm$1.25 & broadened e$^{+}$e$^{-}$ line  \\
\hline
\rule[-1ex]{0pt}{3.5ex} continuum & power law index 2.17 \\
\hline 
\end{tabular}
\end{center}
\caption{Astrophysical nuclear line emissions and continuum source simulated with the combined-imaging telescope.}
\end{table}

\subsection{Energy Resolution}
The telescope energy resolution depends upon the energy resolutions of both the silicon tracker and the CdZnTe absorbers. For each hit, the energy resolution adds in quadrature to get the total energy resolution per event. Eq.~\ref{eq:quadrature} describes the total energy resolution for an \textit{n}-site event, where \textit{E$_{i}$} is the energy resolution of the detector where hit \textit{i} occurred. Because most of the Compton interactions occur in the tracker, the good energy resolution of the Si DSSD detectors is reflected in the overall energy resolution of the telescope.

\begin{figure*}[ht!]
\centering
\includegraphics[width=0.75\linewidth]{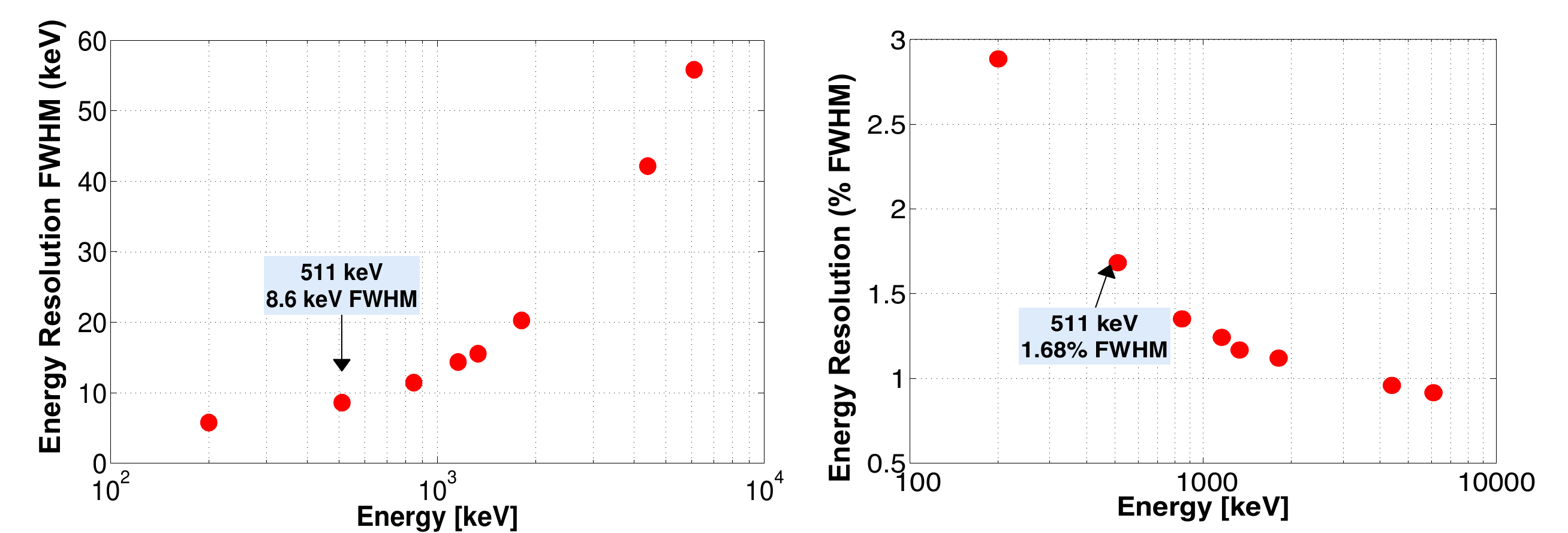}
\caption 
{\label{fig:energyres}Energy resolution, including multiple-site events, as a function of energy in terms of the FWHM, left, and percent, right.}
\end{figure*}

Using the performance parameters described in the previous section, the simulated photopeak energy resolution for the telescope is shown in Fig.~\ref{fig:energyres} as a function of energy in terms of the FWHM, left, and percent, right. This estimation includes multiple-site events. The overall energy resolution of the telescope corresponds to 1.68$\%$ FWHM at 511 keV.

\begin{center}
\begin {equation}
\Delta E_{tot} = \sqrt {\sum\limits_{i=1}^n \Delta E_{i}^{2}}
\label{eq:quadrature}
\end {equation}
\end{center}

\begin{figure*}[ht!]
\centering
\includegraphics[width=0.75\linewidth]{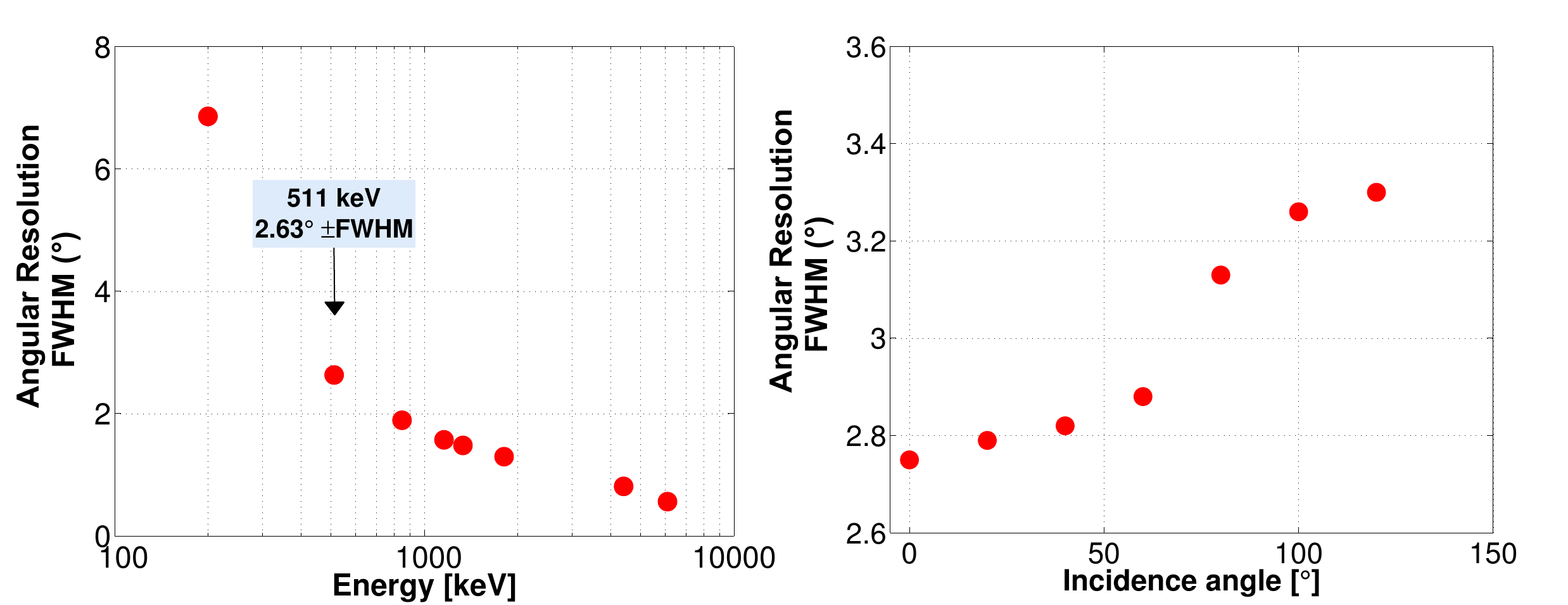}
\caption
{\label{fig:armTelescope}ARM (Compton mode) as a function of energy, left, and incidence angle, right, for a 511 keV broadened line source, right.}
\end{figure*}

\subsection{Angular Resolution}
For Compton events, the ARM is defined as the angular distance between a known source position and the closest reconstructed position on the Compton cone, i.e.\ the ARM is the width of the Compton cone, thus represents the power to resolve point sources (Schoenfelder et al. 1993). It is a function of energy resolution, position resolution, and Doppler broadening. Figure~\ref{fig:armTelescope} shows the ARM as a function of energy, left, and incidence angle for a 511 keV broadened line source, right. A $\pm$3$\sigma$ photopeak energy window was used to determine the ARM. 

The degradation in the ARM at higher incidence angles is due to geometric effects. Because a photon is more likely to scatter into a neighboring detector rather than between planes at higher incidence angles, the average distance between the first two interactions as seen from simulations is shorter, i.e.\ 9.9 cm at 120$^{\circ}$ vs 13.2 cm for on-axis photons. The limiting factor in the achievable angular resolution for the telescope in Compton mode is the position resolution from the 0.5 mm strip pitch of the silicon detectors and particularly the 1 cubic-centimeter voxel size of the CdZnTe detectors. 

The angular resolution of a coded-mask instrument is the smallest angular distance at which 2 point-like sources can be resolved. Conservatively, this distance is taken to be 1.26 times the FWHM of the detected point spread function (Dean $\&$ Byard 1991). The angular resolution is determined by the instrument geometry as shown in Eq.~\ref{eq:mask_angres_skinner} (Skinner 2008), where \textit{m} is the horizontal spacing between the mask pixels, \textit{d} is the position resolution of the detector elements, and  \textit{L} is the distance from the mask to the detection plane. 

\begin{equation}
\delta\theta^{2} =(m/L)^{2} + (d/L)^{2}
\label{eq:mask_angres_skinner}
\end{equation}

Given that the first Compton event occurs in the DSSD, the pixel spacing is much larger than the detector spatial resolution, \textit{d $<<$ m}, thus the angular resolution can be approximated by Eq.~\ref{eq:mask_angres}. 
 
 \begin{equation}
\Theta =arctan(m/L)
\label{eq:mask_angres}
\end{equation}

The mask-detector distance is a trade-off between Compton efficiency, as decreasing the separation distance between planes yields more Compton scatter events, and angular resolution, as larger separation distances yield better angular resolution in both Compton and coded-mask mode. Furthermore, at larger standoff distances the background is minimized, since gamma rays that originate from activation in mask materials are less likely to hit the detection plane. A separation distance between the mask and the top plane of the tracker of {\em L} = 2.3 m and a pixel spacing of {\em m}= 0.5 cm were chosen to yield an angular resolution of 0.125$^{\circ}$ within a 10$^{\circ}$ FoV and at all energies. This is roughly a factor of 20 improvement at 511 keV over the achievable angular resolution in Compton mode.
 
In the case where \textit{d $<<$ m}, the degree of uncertainty to which a point source can be located, the Point Source Location Accuracy (PSLA), is given by Eq.~\ref{eq:mask_psla_Stephen} (Stephen 1991), where $\sigma$ is the significance of detection. Thus a 3$\sigma$ detection yields a position uncertainty of $\pm0.04^{\circ}$. 

\begin{equation}
PSLA(1\sigma) =arctan(\frac{m}{L})*\frac{1}{\sigma}
\label{eq:mask_psla_Stephen}
\end{equation}

For image reconstruction, the first hit can occur in any detection plane of the tracker, thus the angular resolution improves slightly as a function of interaction depth and is also energy dependent. However, at 511 keV, the probability that the first Compton scatter occurs in the top half of the tracker is $>$70$\%$. In this case the improvement in angular resolution is less than $\sim$4$\%$, thus is considered negligible for this study.

\subsection{Effective Area}
\begin{figure}[ht!]
\centering
\includegraphics*[width=0.75\linewidth]{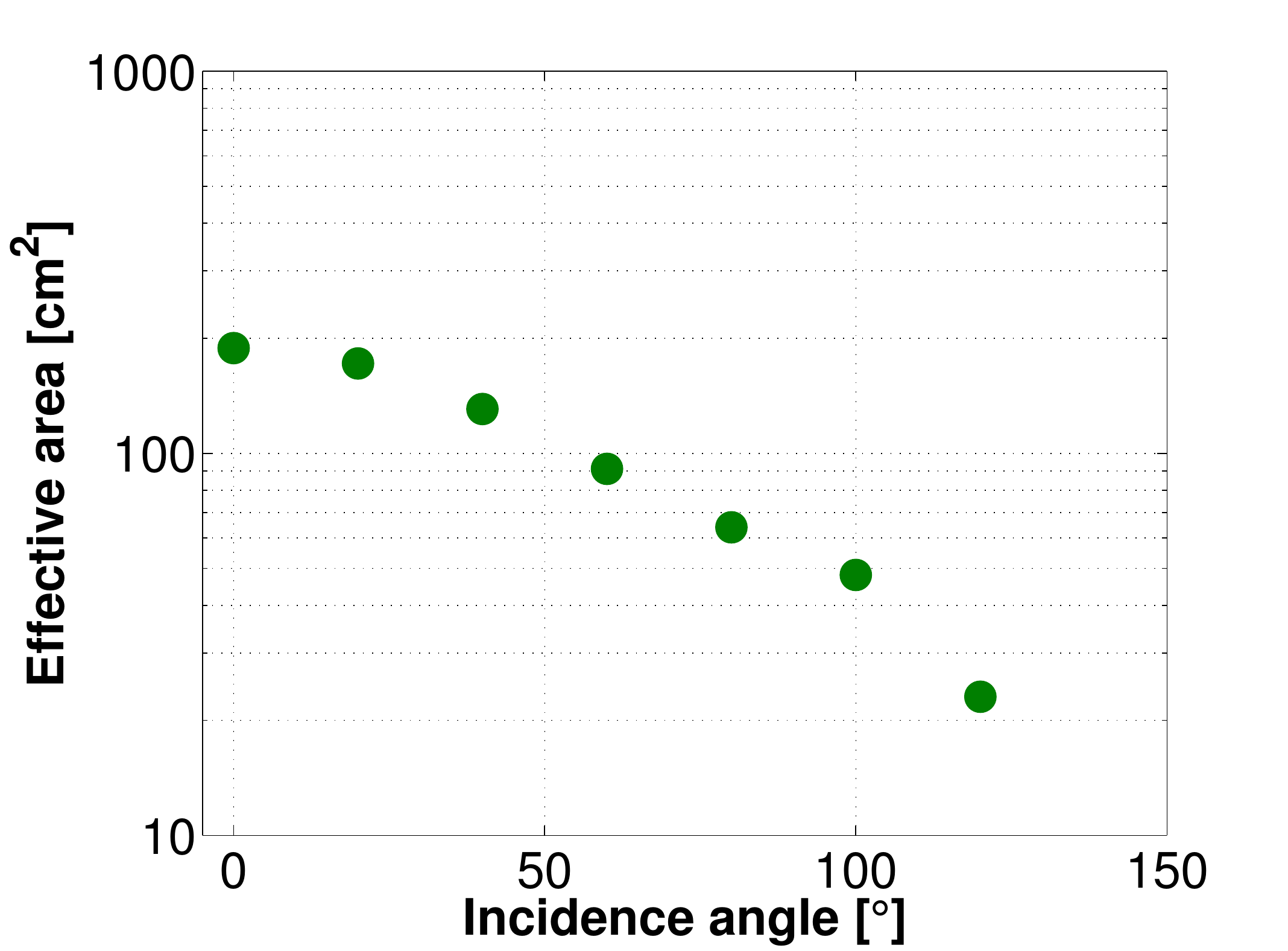}
\caption{Effective area as a function of energy, top, and as a function of incidence angle for a broadened 511 keV source, bottom.}
\label{fig:Aeff_energy_angle}
\end{figure}

Because not all of the photons that are incident upon the geometric area of a detector interact to produce counts or real events in the detector, the effective area is a reduction from the actual geometric area of a detector. This reduction is due to the intrinsic detection efficiency and the event cuts required for Compton reconstruction. The event cuts include an energy window of $\pm$1.4$\sigma$ around each line energy, an ARM cut of $\pm$3$\sigma$, and an Earth horizon cut. The effective area as a function of energy and of incidence angle is shown in Fig.~\ref{fig:Aeff_energy_angle}, e.g. the effective area is 182.7 $\pm$0.4 cm$^{2}$ for the Compton telescope without the mask at 511 keV. As the characterization is based on simulation, the uncertainties are statistical and come from the number of simulated events. No mask was included in these simulations.

\begin{figure}[ht!]
\centering
\includegraphics*[width=0.75\linewidth]{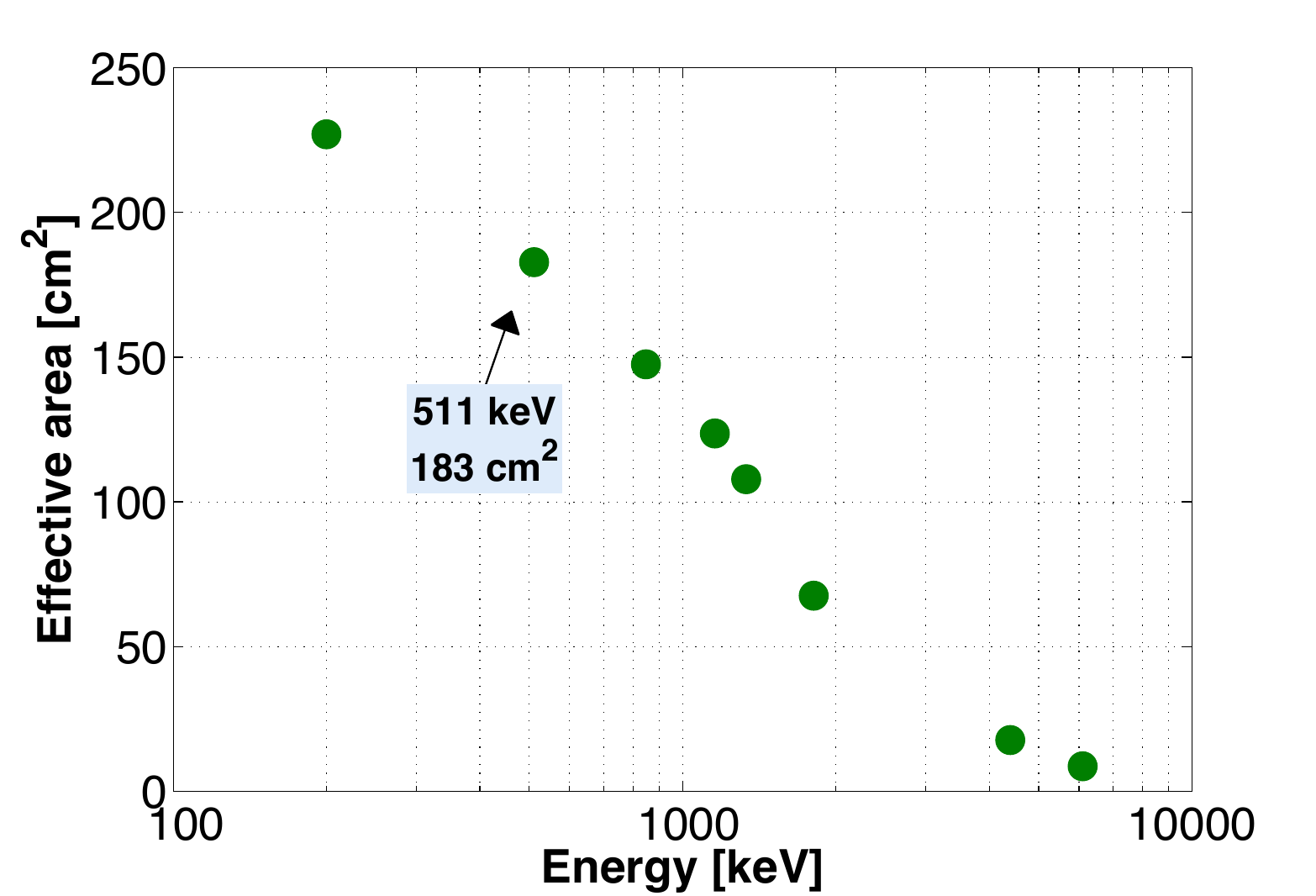}
\end{figure}

\section{Sensitivity Estimates}
The sensitivity of a Compton telescope can be determined based upon the effective area estimates and a model of the expected background flux. There are several challenges to achieving a high sensitivity in a Compton telescope. The flux from astrophysical gamma-ray sources is typically low, especially compared to the background flux from cosmic rays and the Earth's albedo. Additionally activation of both passive and active components of the telescope, particularly from cosmic-ray interactions as well as charged particles trapped in the Earth's magnetic fields, creates a pervasive source of background which generally either reaches a statistical equilibrium or continues to increase over time. In addition to the low flux from target gamma-ray sources, the total interaction cross-section reaches a minimum in the Compton regime. Therefore good background reduction techniques are essential for obtaining high sensitivity and can be achieved through a combination of a good telescope design, good energy and angular resolution, CSR discrimination, and through careful event selections.

Additional challenges are present for achieving high sensitivity in coded-aperture mode. Because source localization is deduced by projecting the spatial locations of the detected counts onto the sky, the background counts are also backprojected along with source counts, thus complicating the ability to discriminate source from background in mask mode. An effective deconvolution algorithm or the ability to operate the telescope in on/off mode to distinguish source from background can allow for better discrimination. These techniques are most effective under the assumption of a uniform background distribution, however can be applied when long exposure times allow for accurate background characterization. 

In the case of a passive coded mask, events are lost due to absorption in the mask, thus reducing the already limited number of source photons. Additionally, the mask in most cases is either not perfectly transparent nor perfectly opaque. Thus the loss is a function of both the filled area of the mask, the energy dependent transmission through the mask, as well as absorption or scattering in materials that surround the sensitive detector volume, such as support structures or dead layers in the detectors.  

In addition to these nonidealities, the sensitivity and effectiveness of mask-mode observation depends also on the angle of incidence of the source. With regards to this design, only a partially coded field-of-view can be obtained for all incident sources that are outside of the 10$^{\circ}$ pointing mode.

\subsection{Background Model}
Because gamma-rays are absorbed by the Earth's atmosphere, observations of astrophysical sources require placing the telescope at balloon altitudes or in space. The expected background is hugely dependent upon the choice of orbit for the satellite. In this case, a near-equatorial, low-altitude orbit was chosen. This orbit would minimize exposure to cosmic rays and to high-energy charged particles in the South Atlantic Anomaly. A 575 km orbit with a 6$^{\circ}$ inclination has been extensively studied for the NuSTAR mission (NuSTAR collaboration), thus the NuSTAR pre-flight background model was used for this study. 

\begin{figure}[ht]
\centering
\includegraphics*[width=1.0\linewidth]{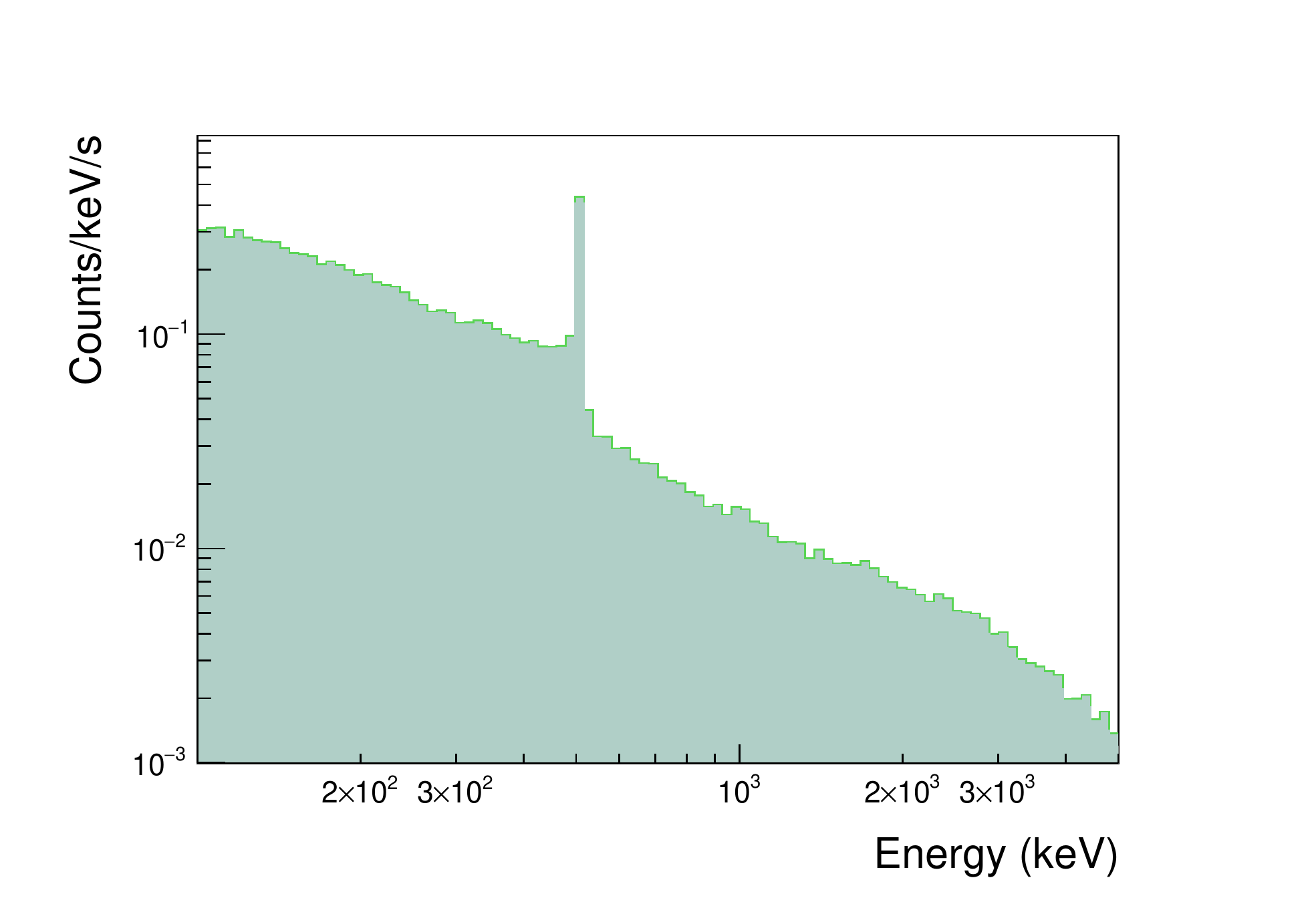}
\caption{\label{fig:bkgrd575}Reconstructed background after sensitivity optimized event cuts for a 575 km orbit with a 6$^{\circ}$ inclination. Simulated background includes cosmic and albedo photons, leptons, and hadronic components from both prompt interactions and activation.}
\end{figure}

The gamma-ray background, shown in Fig.~\ref{fig:bkgrd575}, consists of extragalactic  X-rays and gamma rays, annihilation photons, albedo photons up to several MeV, electrons, positrons, protons, and neutrons. Activation within passive and active volumes of the instrument from hadronic interactions were included in the simulation. The particle distributions were isotropic except for the observed enhancement of albedo photons at zenith angles of 120$^{\circ}$. The Compton reconstructed background components were simulated and then reconstructed using the BackgroundMixer program in MEGAlib. From the multiple components one can see that the 511 keV background is excessive, thus making high sensitivity of true sources at this energy very challenging.

\subsection{Sensitivity in Compton Mode}

The telescope sensitivity is a description of the minimum source flux which can be measured by the instrument within a given significance above background. At the sensitivity limit, the number of detectable source counts, {\em $N_{s}$}, is given by the minimum flux, {\em F$_{min}$}, the effective area, {\em $A_{eff},$} and the observation time, {\em $T_{obs}$}, according to Eq.~\ref{eq:sourcecounts}.

\begin{equation}
N_{s} =F_{min}A_{eff}T_{obs}
\label{eq:sourcecounts}
\end{equation}

The number of source counts that satisfies the minimum detectable flux is dependent upon the signal-to-noise ratio, $\sigma$, as shown in Eq.~\ref{eq:snr}, where {\em $\Delta$M$_{s}$} is the uncertainty in the measurement.

\begin{equation}
{\sigma} = \frac{N_{s}}{{\Delta}M_{s}}
\label{eq:snr}
\end{equation}

Under the condition of Gaussian or Poisson statistics, the measurement uncertainty is described by Eq.~\ref{eq:measuncertainty}, which considers the uncertainty in measuring both the source counts, {\em $\Delta$S}, and the background counts, {\em $\Delta$B}, where N$_{b}$ is the number of detected background counts.
\begin{equation}
\Delta M_{s} = \sqrt{ \Delta S^{2} + \Delta B^{2}} = \sqrt{N_{s} + N_{b}}
\label{eq:measuncertainty}
\end{equation}

An expression for the minimum flux, i.e. sensitivity of a telescope for a given source energy, can be derived using Eqs.~\ref{eq:sourcecounts} and ~\ref{eq:snr}. In this case $\sigma$ is a factor of {\em n}, defined as the detection significance of the source above background, typically 3$\sigma$ or 5$\sigma$. The resulting equation is shown in Eq.~\ref{eq:minflux} (Jacobson 1975).

\begin {equation}
F_{min}(E) = \frac {n^{2} + n\sqrt{n^{2} + 4N_{b}}}{2A_{eff}T_{obs}}
\label{eq:minflux}
\end {equation}

\begin{figure}[ht]
\centering
\includegraphics*[width=0.75\linewidth]{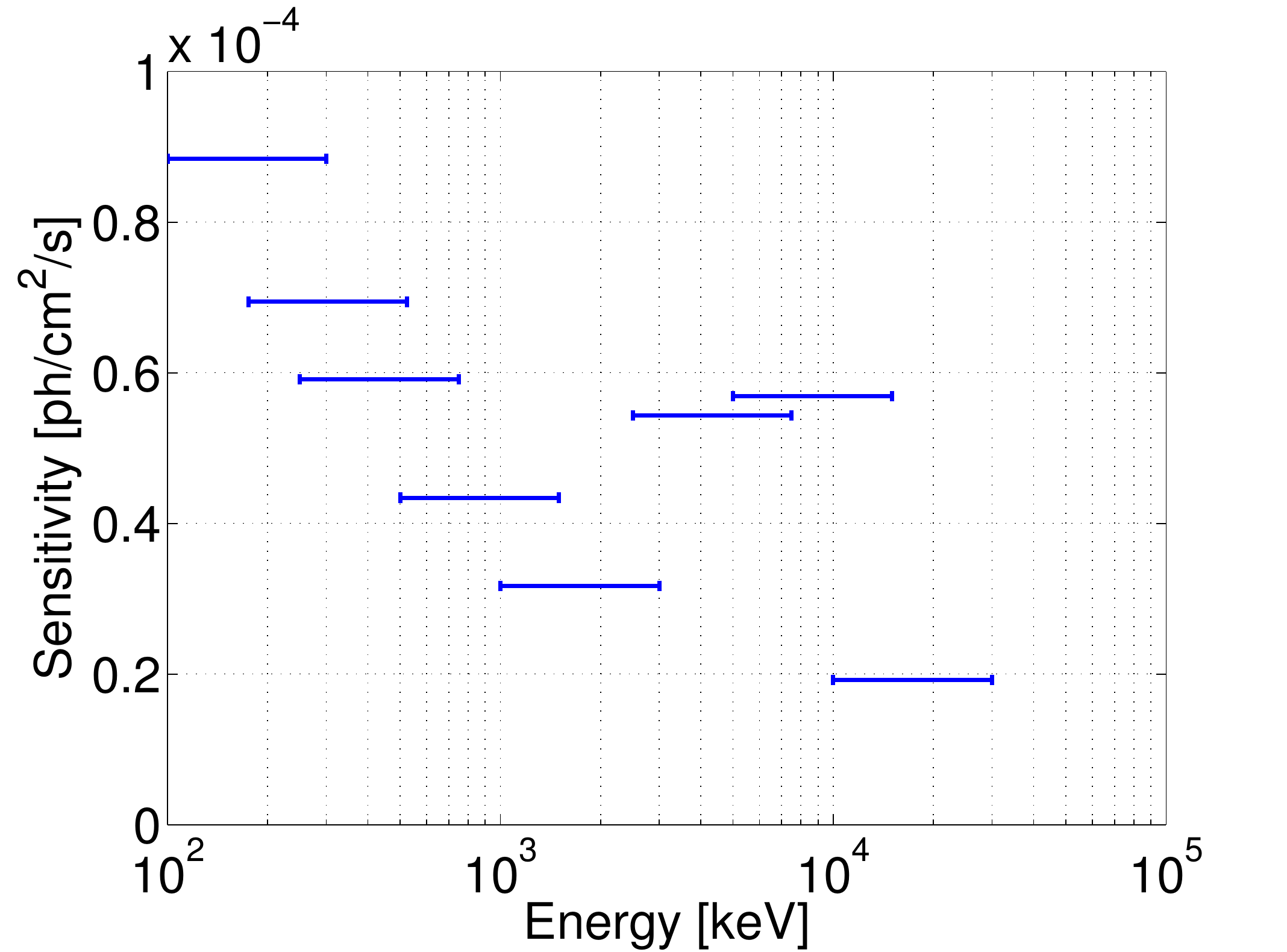}
\caption
{\label{fig:sensitivity_coco}3$\sigma$ sensitivity as a function of line source energy, top, and continuum sensitivity for a simulated Crab-like source, bottom, over an observation time of 1 Ms.}
\end{figure}

Figure~\ref{fig:sensitivity_coco} shows the calculated 3$\sigma$ sensitivities with optimized event cuts as a function of energy for monoenergetic narrow line sources, left and the continuum sensitivity for a simulated Crab-like source with a power law index of  2.17, right. The observation time was 1 megasecond in pointing mode. For scanning the whole field-of-view, a 511 keV broadened line source was simulated separately, resulting in an all-sky sensitivity of \mbox{3.6 x 10$^{-6}$ ph cm$^{-2}$ s$^{-1}$} over a two-year observation time. Simulations for the calculated sensitivities do not include the coded mask.

\begin{figure}[ht]
\centering
\includegraphics*[width=0.75\linewidth]{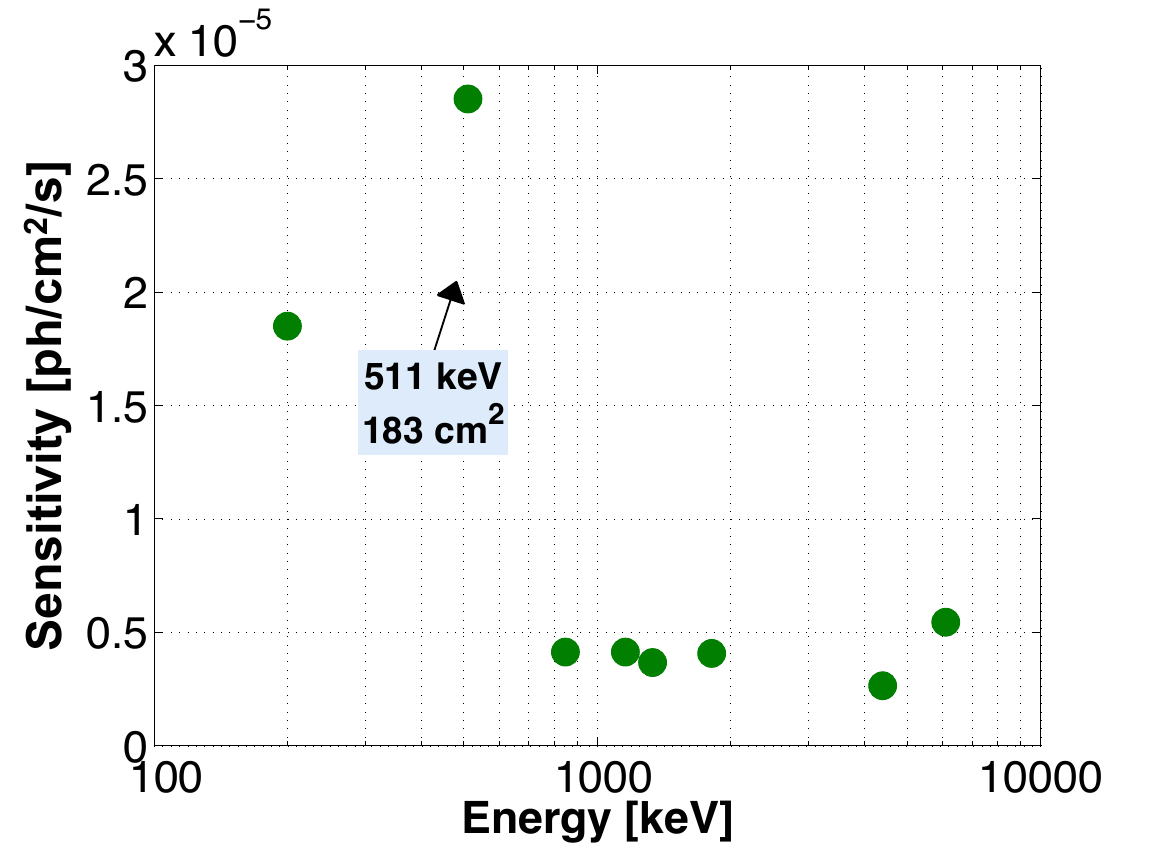}
\end{figure}

\subsection{Sensitivity in Combined Compton and Coded Mask Mode}

To evaluate the sensitivity of a coded-aperture instrument, several factors that can contribute to image non-uniformity and that are also relevant for Compton-coded-mask (CCM) imaging mode need to be considered (Skinner 2008). For this study, no support structure exists for the mask, the simulated detector response is identical for each element, the observation times allow for Gaussian statistics, and the modeled background is spatially uniform. The sensitivity in mask mode is evaluated at or near the center of the field-of-view, thus considering only fully-coded detection scenarios and neglecting non-idealities due to the thickness of the mask elements. Furthermore, the mask sensitivity is evaluated only when combined with the Compton telescope. In this case, source photons that pass through the mask unimpeded due to the energy-dependent opacity of the mask elements are compensated for by their deposition in the detection planes and subsequent Compton reconstruction. This allows for background reduction before the mask elements are used to enhance the angular resolution.

As consequence, the sensitivity in CCM mode can be calculated using the same equation as in Compton mode, as the background events contributing to the CCM sensitivity are the same events that contribute to the source in Compton mode. Under the above scenario and considering a mask with open fraction \textit{f} = 1/2, the number of source counts (i.~e.~ the effective area) is reduced by half, thus requiring double the minimum flux as compared to Compton-mode only. This approximation is valid within the fully-coded field-of-view, assuming only single sources are present, and considering a background-dominated case where detection occurs at the sensitivity limit.

\begin{figure*}[ht!]
\centering
\includegraphics[width=0.7\linewidth]{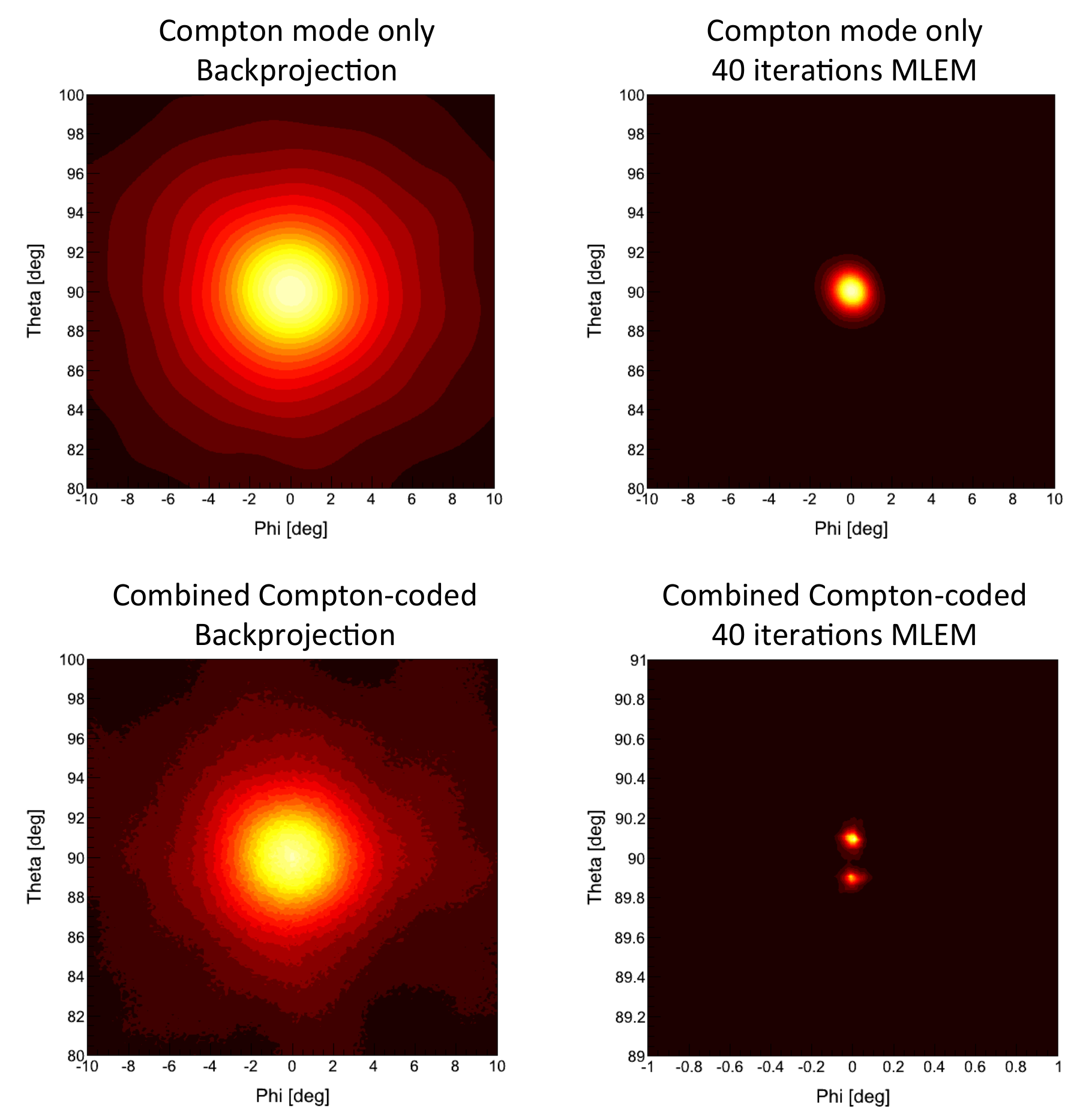}
\caption
{\label{fig:CoCoImages}Imaging of two sources separated by 0.2 degrees. Top left: Back projection in Compton mode only. Bottom left: Back projection with combined Compton-coded mask mode. Top right: Deconvolved image after 40 iterations in Compton mode only. Bottom right: Combined Compton and coded-mask mode image after 40 iterations. Note the change in scale between the bottom two images.}
\end{figure*}

\section{Image Reconstruction}

Figure~\ref{fig:CoCoImages} shows a demonstration of the multimode imaging capabilities of the telescope. Simulations of two 511 keV point sources separated by 0.2 degrees were performed. A list-mode maximum-likelihood expectation-maximization algorithm was used for the deconvolved images. In the top two images, only reconstructed Compton events were used. Both before and after deconvolution the two sources are not separable. The image reconstruction in the bottom two images takes into account the absorption probabilities of the initial photons through the coded mask. The image before the deconvolution shows some additional noise due to the overlapping coded-mask patterns superimposed onto the Compton imaging space. However, the deconvolved image clearly shows that the two sources can be separated.


\section{Comparison with Other Space Instruments}
To assess the capabilities of the CdZnTe-based combined imager as a future telescope, Table~\ref{tab:telescopes511} shows a comparison of performance parameters for a broadened 511 keV source of the instrument to other medium-energy gamma-ray telescopes, e.g., the coded mask telescope INTEGRAL/SPI and the proposed e-ASTROGAM Compton telescope (De Angelis 2017). The latter is based upon a silicon tracker and a thallium-activated cesium iodide calorimeter. One can see from the table that for nuclear line spectroscopy, the germanium-based SPI instrument significantly exceeds the performance of the CdZnTe-based telescope in terms of energy resolution. However, the angular resolution without the mask is comparable to the other two instruments. A big improvement at lower energies is seen in angular resolution with the use of the mask. Theoretically, a mask can be added to any future Compton telescope in order to achieve a high angular resolution at energies below $\sim$0.5 MeV, depending upon the mask thickness and pixel material. 

In terms of sensitivity, the CdZnTe-based combined imager is a good candidate for a next-generation telescope. In both imaging modalities, it exceeds the sensitivity of SPI and, within the fully-coded field-of-view, is comparable to that of \mbox{e-ASTROGAM} at 511 keV. However, a direct comparison is difficult because the geometric area of \mbox{e-ASTROGAM} and consequently its effective area are much larger than for this telescope. For a direct comparison, the instrument would need to be scaled and other factors would also need to be considered, such as cost and detector availability.

\begin{table*}[ht!]
\begin{center}       
\begin{tabular}{rcccc} 
\hline\hline
\rule[-1ex]{0pt}{3.5ex}      & \textbf{This Work} & \textbf{INTEGRAL/SPI} & \textbf{e-ASTROGAM} &   \\
\hline
\rule[-1ex]{0pt}{3.5ex}    Energy Resolution   & 1.8   &  0.38   &  3  &   \\
($\%$ FWHM) \\
\hline
\rule[-1ex]{0pt}{3.5ex}      Angular Resolution         & 2.1$^{\circ}$ Compton & 2.5$^{\circ}$  & $2.5^{\circ}$ &  \\
\rule[-1ex]{0pt}{3.5ex}     ($^{\circ}$ FWHM) &  $<$0.13$^{\circ}$ with mask  \\
\hline
\rule[-1ex]{0pt}{3.5ex}      Effective Area	& 143   & $\sim$100 & 446  &\\
       (cm$^{2}$)   &  $\sim$72 with mask\\
\hline
\rule[-1ex]{0pt}{3.5ex}      3$\sigma$ Sensitivity on-axis &  2.9 x 10$^{-5}$ & 5.2 x 10$^{-5}$  & 4.1 x 10$^{-6}$  &\\
      (ph cm$^{-2}$ s$^{-1}$)	&  5.8 x 10$^{-5}$ with mask  \\
\hline 
\end{tabular}
\end{center}
\caption{Telescope performance comparison of a broadened line source at 511 $\pm$ 1.3 keV for a $10^{6}$ second observation time. For this study and for e-ASTROGAM, the angular resolution was optimized for sensitivity, and reconstruction was performed without electron tracking.}
\label{tab:telescopes511}
\hspace{40cm}
\end{table*}

\par


\section{Summary}
Based upon laboratory measurements and benchmarked simulations of the HEMI CdZnTe detectors, the combined imager studied in this work has achievable energy resolutions of 1.68\% FWHM at 511 keV and 1.11\% at 1809 keV, on-axis angular resolutions in Compton mode of  2.63$\,^{\circ}\pm$ FWHM at 511 keV and 1.30$\,^{\circ}\pm$ FWHM at 1809 keV, and is capable of resolving sources to at least 0.2$\,^{\circ}$ at lower energies, e.g., 511 keV, with the use of a coded mask. An initial assessment of the instrument yields an effective area of 183 cm$^{2}$ at 511 keV and an anticipated all-sky sensitivity of \mbox{3.6 x 10$^{-6}$ photons cm$^{-2}$ s$^{-1}$} for a broadened 511 keV source over a 2 year observation time. Additionally, combining a coded mask with a Compton imager to improve point source localization for positron detection has been demonstrated. 

These capabilities meet several scientific objectives in medium-energy gamma-ray astronomy, such as achieving the sensitivity and accuracy required for nuclear line studies and, in particular, allowing for improved observations of the spatial distribution of Galactic 511 keV emission. Although a future telescope with better energy resolution would significantly improve the sensitivity to line emission, the achievable sensitivity from this design would serve well as a telescope for an all-sky survey. A CdZnTe detector array, in particular, is a good candidate as an absorption plane behind a tracker, as demonstrated, or as an absorbing focal plane behind a wave optics telescope, such as in a Laue lens instrument.

Although this study focused primarily on applying the HEMI detector technology and multimode concept to a space mission, improvements to the simulated performance are anticipated with the use of the silicon tracker in its full capacity. The tracker is capable of reconstructing the direction of the recoil electron for incident photon energies above $\sim$2 MeV. This reduces the reconstructed Compton circle to an arc, as many incident directions can be eliminated by tracking the recoil electron. Additionally, for energies higher than 5-10 MeV pair production processes dominate in silicon, thus it can be used to track the electron-positron and determine the initial photon direction using conservation of momentum.


\clearpage
\newpage
\twocolumn

\end{document}